\documentclass{article}

\usepackage{aimc2025}
\usepackage{url}
\usepackage{enumitem}
\setlist[enumerate]{noitemsep} 
\usepackage{verse}
\usepackage[utf8]{inputenc} 
\usepackage[T1]{fontenc}    
\usepackage{hyperref}       
\hypersetup{
  colorlinks   = true, 
  urlcolor     = blue, 
  linkcolor    = black, 
  citecolor   = black 
}
\usepackage{url}            
\usepackage{booktabs}       
\usepackage{amsfonts}       
\usepackage{nicefrac}       
\usepackage{microtype}      
\usepackage{graphicx}       
\usepackage[normalem]{ulem}

\title{``I made this (sort of)'': Negotiating authorship, confronting fraudulence, 
and exploring new musical spaces 
with prompt-based AI music generation}

\author{%
  Bob L. T. Sturm \\ 
  Division of Speech, Music and Hearing (TMH)\\
  KTH Royal Institute of Technology\\
  Stockholm, Sweden 10044 \\
  \texttt{bobs@kth.se} 
}

\begin{document}

\maketitle

\begin{abstract}
I reflect on my experience creating two music albums 
centered on state-of-the-art prompt-based AI music generation platforms.
The first album explicitly poses the question:
What happens when I collide my junk mail with these platforms?
The second album is a direct response to the first,
and toys with the inability of 
state-of-the-art prompt-based AI music generation platforms
to generate music that is not ``practiced'', ``polished'', and ``produced''.
I seed a large language model (LLM) with information about these albums
and have it interview me, which results in the exploration of several deeper questions:
To what extent am I the author? Where am I in the resulting music?
How is my musical identity changing as I am 
faced with machines that are in some ways far more talented than I?
What new musical spaces does my work open,
for me or anyone/thing else?
I conclude by reflecting on my reflections, 
as well as LLM-mediated self-reflection as method.
\end{abstract}

\section{Introduction}
I released two music albums in late 2024, each
wrestling with state-of-the-art prompt-based AI music audio generation platforms,
specifically Suno\footnote{\url{https://suno.com}, last accessed July 30 2025.} and Udio.\footnote{\url{https://www.udio.com}, last accessed July 30 2025.}
I built the first album, {\em Music from the Spam Folder} \citep{Sturm2024b},\footnote{\url{https://bobltsturm.bandcamp.com/album/music-from-the-spam-folder}, last accessed July 30 2025.}
by ``colliding'' these platforms with junk mail (spam) I have received.
More specifically, I fashion prompts from spam, 
input these to AI music generation platforms,
curate audio recordings from their outputs, and then build tracks
using a digital audio workstation (DAW).
(See \cite{Sturm2024b} for details on the process.)
Initially started as a playful way to explore the creative text-setting abilities of these platforms,
deeper consideration of my process, not to mention my astonishment at the quality of the results, 
consolidated into a vision of exploring spam as text---an experimental 
``musicalization of everyday life'' \citep{Tan2024a}---as well as experiencing the creation and release of an album of 
AI-generated (or at least co-created) music.

The second album, {\em A Difficult Christmas} \citep{Sturm2024c},\footnote{\url{https://bobltsturm.bandcamp.com/album/a-difficult-christmas}, last accessed July 30 2025.}
is a direct response to the other.
Whereas {\em Music from the Spam Folder} marvels at
the disconcerting musical creativity of 
state-of-the-art prompt-based AI music audio generation platforms,
{\em A Difficult Christmas} revels in what I have discovered through trial and error
they seem unable to do: generate anything that is not practiced, polished, and produced---
probably a direct result of their being trained on commercially released professionally recorded music.
What I mean by {\em practiced} is that the musicians heard in a recording
have mastery over their instruments and the music they are playing,
and someone has approved their recorded performance to be released for sale.
By {\em polished} I mean that aspects of a recorded performance
have been ``touched up'', e.g., particular takes have been 
selected and combined with other takes, pitches have been corrected,
and so on.
And what I mean by {\em produced} is that the polished recordings have been
further processed, e.g., reverberated, equalized, and mixed down to stereo with dynamic compression,
to sound good on a variety of sound platforms.
{\em A Difficult Christmas} begins with a four-minute track composed from
prompts rendered as speech and outputs generated by Udio from them.
Each prompt is an attempt to make Udio generate something
that sounds like a novice violin student practicing.
The following ten tracks feature myself playing solo violin by ear without any practice
or mastery of the instrument, emitting kinds of musical sounds and expression
that at this time are not possible with Udio---and probably 
will remain so until my album finds its way into its training data.
The album embraces imperfection as a form of authenticity,
as in Charles Ives' bandmaster father responding to a complaint of 
out-of-tune singing: ``Don’t pay too much attention to the sounds---for if you do, you may miss the music''
\citep{Hentoff1974a}.

This next section of this paper consists of self-reflection about these albums,
mediated by a large language model (LLM) 
prompted with the liner notes of the albums, a version of the abstract and introduction above, 
and the following text:
\begin{quote}
I would like to be interviewed about two albums I have produced (liner notes attached). This will be edited and submitted to the 2025 AI Music Creativity conference as a paper. AIMC this year explores the theme, "The Artist in The Loop", which invites innovation and analysis of the many forms of feedback loops and the many touch points between creators and AI platforms across timescales and human scales. Of particular interest are: Search and Discovery: how can AI enable artists to explore spaces of creative possibilities and discover points of interest, even profound novelty? Control and Differentiation: how can we overcome problems of control of generative AI platforms, and how does an artist reinsert their distinctive selves into processes of AI creation? Rich Engagement Beyond the Output: musical output is not everything. How can AI support musical creation that has expanded benefits – enriching understanding of music theory, music history, production techniques, nurturing social or cultural expression, or integrating music into rich multidimensional creative contexts. Other relevant conference topics include: New AI tools for composition or performance; Formal analysis of music AI platforms; Creative practice-based research in music AI; New applications of music AI; Understanding human, machine and collective creativity in music AI; HCI for music AI; Theory, concepts, frameworks relating to music AI practice; Research into the global cultures of music AI; and Ethics of music AI.

Ask me one question at a time about my work, addressing the themes of AIMC. Also refer to the literature where possible with proper citations. You may also create illustrations to make your point.
\end{quote}
The LLM then prompted me with some text upon which to reflect. 
I then drafted a response and input this back to the LLM,
which then responded with a new prompt.
I proceeded in this way until this manuscript reached 5,000 words 
(the maximum length of conference paper submissions to AIMC 2025).
I finally revisited and edited all my responses, the abstract and introduction,
and wrote a final section reflecting on my self-reflections and this method.\footnote{This paper
was not accepted to AIMC 2025.}

\section{LLM-mediated Self-reflection}
\subsection{On Authorship and the Distinctive Self}
{\em ChatGPT 4o}: {\tt In the liner notes of ``Music from the Spam Folder'', you describe feeling ``wary of taking too much credit'' \citep{Sturm2024b} 
and liken your role to that of someone making a mixtape. Given AIMC’s theme of “The Artist in the Loop”, how do you understand your authorship in this project? Do you feel you reinsert your distinctive self into the generative process --- and if so, how?}

In the context of prompt-based music audio generation platforms like Suno and Udio,
questions of authorship, fraudulence, and idiosyncrasy 
are ones with which I have been wrestling since late 2023.
How is anyone the author of the music in an audio recording generated from 
whatever prompt they have written, no matter how original it may be?
They are as much an author as someone ordering food at a restaurant is a chef.
Similar questions have been asked for many years in the context of
computer generated art \citep{McCormack2019a},
and recently in the context of generative AI \citep{Caldwell2023a,Anscomb2025a}.
The output of these platforms is often a finished product
that does not present many creative opportunities---so the term ``reinsert'' is appropriate.
For me, it is not easy to find a way to contribute enough such that
I feel like an author.
Other than the specification of some platform parameters, 
the only affordance of Suno and Udio is text input (and recently audio excerpts);
and while prompting seems to provide innumerable possibilities, 
a user has only three other avenues for making music with these platforms: 
curating, covering/composing, and (re)mixing/sampling.

{\em Prompting} is the crafting of material used to condition a generative model,
e.g., the text ``A medium-tempo funky song about music informatics researchers 
meeting in Germany with Irish-style melodic lines played by bass,
drums, acoustic guitar, piano, vocoder, baritone sax …'' 
Being able to prompt generative models with language
takes advantage of a wide-spread skill (literacy),
and so has become a very popular interaction paradigm \citep{Li2024a}---though some consider it ``harmful'' and a ``fad'' \citep{Morris2024a}.
The skills and strategies behind ``prompt engineering''   
are being recognized and studied \citep{Dang2022a,Zamfirescu-Pereira2023a},
and some are proclaiming: ``We have entered an era in which anybody can generate digital images from text---a democratization of art and creative production'' \citep{Oppenlaender2024a}.
Some are protective of their prompts as if they are intellectual property \citep{Yao2024a},
which motivates the buying and selling of prompts.\footnote{For example, see PromptBase \url{https://promptbase.com}, last accessed July 30 2025.}
Will this result in a further widening of the concept of ``musical property'' \citep{Scherzinger2014a}?
An irony here is that such possessiveness is of something used 
to make a generative AI platform mimic the idiosyncrasies of 
the authors of the content on which it has been trained---itself often acquired and used without explicit permission,
e.g., as alleged against Suno and Udio \citep{Suno_suit2024,Udio_suit2024}.
Furthermore, one need only copy and paste a prompt to duplicate the output ``vibe''
of the original prompt author.\footnote{This alludes to Andrej Karpathy's term ``vibecoding'', 
which is prompting a large language model to generate computer code 
meeting described specifications \url{https://x.com/karpathy/status/1886192184808149383}, last accessed July 30 2025.}
Someone asserting authorship over the output of Suno and Udio simply because 
they created the prompt, whether by spending many credits experimenting
or simply copy and pasting, feels rather weak.

The second avenue for creating music with Suno and Udio is {\em curating},
which is the selection of generated outputs based on a desire to engage with them.
This could include having the platform extend or edit its output one way or another, 
possibly adjusting the prompt,
or downloading the output for use elsewhere \citep{Tan2024a}.
In curating \citep{Lubar2014a}, one applies editorial considerations,
which include objective judgements, e.g.,
the sung lyrics are clear or the sound quality is high,
and subjective judgements,
e.g., I like the rhythm or I like the mood.
This is where the notion of creating a mixtape comes in:
one curates tracks and compiles them together for one reason or another, e.g., to tell a story.
Through doing so one can impinge on the collection
characteristics of their voice;
but still it seems a far stretch to claim
they are then an author of any of the collected works.
The title ``editor'' feels more appropriate.

The third avenue for creating music with Suno and Udio
is {\em covering/composing}: rendering generated output
with human performance in real life, 
adhering to the original or merely adopting from it.
An example of covering is ``Funk and Data in Deutschland'',\footnote{\url{https://audiolabs-erlangen.de/resources/MIR/2024-DagstuhlThemeSong}, last accessed July 30 2025. \label{funkdata}}
which is a song iteratively generated using Suno and Udio,
and then arranged and performed live by a group of music informatics engineers.
The AI-mediated music making thus moves from presentational to participatory \citep{Turino2008a,Tan2024a}.
An example of composing is ``Cnote'' for tape and double bass by Eric Lyon,\footnote{\url{https://youtu.be/8Elq7tuE5Ng}, last accessed July 30 2025.}
which uses music material generated by Suno.
This avenue can be seen as re-establishing the reality of 
the music audio output by an AI music platform---Baudrillard's ``simulacra'' \citep{Baudrillard1994a,Tan2024a}.
In this case it makes sense to expand the circle of authorship to include oneself.

Finally, the fourth avenue for creating music with Suno and Udio is (re)mixing/sampling.
{\em Mixing} is the processing of digital audio 
to remove artifacts, or make it sound produced, e.g.,
using multiband compressors and reverberation.
{\em Remixing} is where one creates an identifiable version of a piece
through combining its parts (typically voices or stems) in different ways.
{\em Sampling} is a more dramatic act of composition
spanning from the quotation of recorded musical material 
in new contexts, to the complete obliteration of a 
piece of recorded music, e.g., plunderphonics \citep{Oswald1985a,Holm-Hudson1997}.
An example of sampling is ``Fake''
by Phivos-Angelos Kollias,\footnote{\url{https://phivos-angelos-kollias.com/fake}, last accessed July 30 2025.}
which is composed from hundreds of fragments of AI-generated music audio.
Since mixing changes only the surface attributes of a recording,
authorship does not seem warranted.
Remixing and sampling however introduce novelty,
and one's authorship could certainly feel appropriate
depending on the extent of the transformation
of the original material.

\begin{figure}[t]
\centering
\includegraphics[width=1\linewidth]{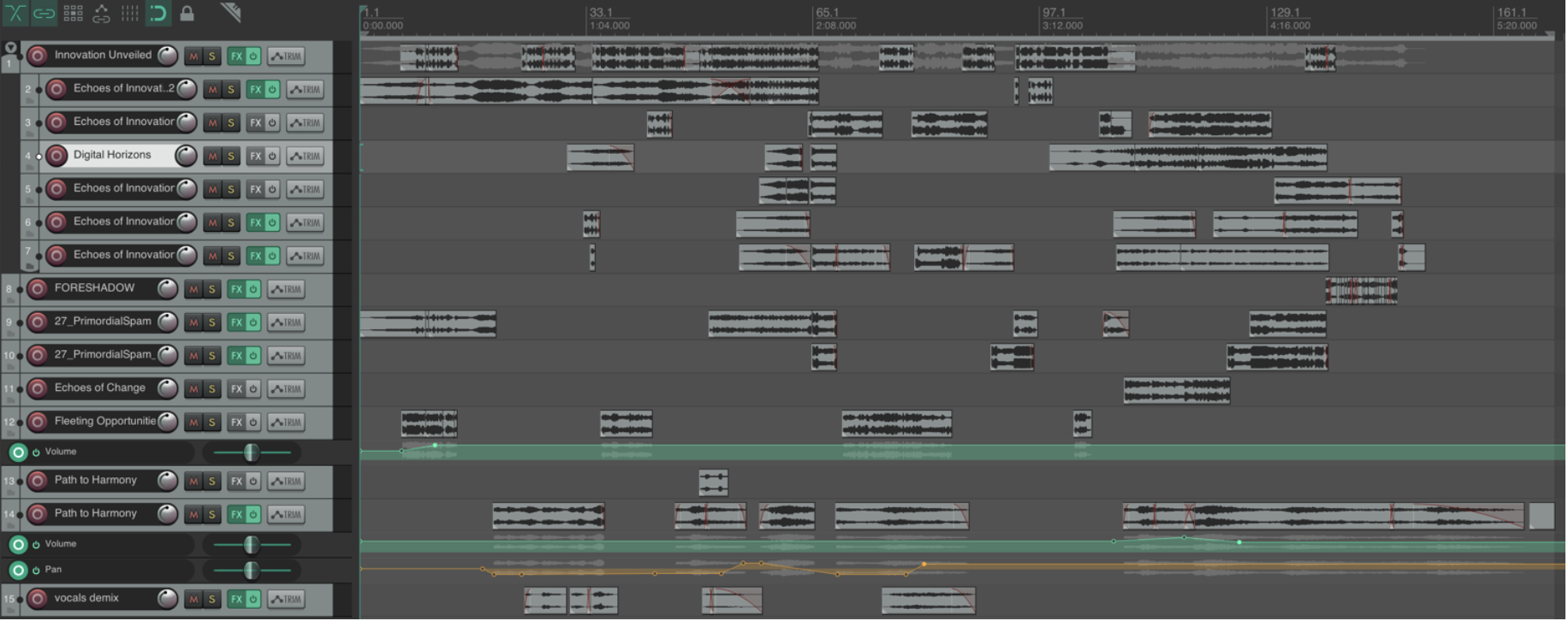}
\caption{The digital audio workstation view of the track ``Peace is the Way'' (track 1) from
``Music from the Spam Folder'' showing how it is composed from different materials
generated by Udio.}
\label{fig:peaceistheway}
\end{figure}

Returning to {\em Music from the Spam Folder},
I see my role as author and as editor spanning different degrees.
The concept of my ``mixtape'' is all mine, 
as is my choice of what spam to work with.
Though I did not author any of the spam, I do render them as prompts,
curate a collection of generated artifacts, and design an exhibit that tells a story \citep{Lubar2014a}.
For some tracks, e.g., ``Fortement en Désaccord, Fortement d'Accord'' (track 7),
and ``Rock Hard!'' (track 13), the generated recordings are presented with very little change;
for others, e.g., ``Peace is the Way'' (track 1) and ``Penny from Automation'' (track 16),
I have performed extensive remixing and sampling.
Figure \ref{fig:peaceistheway} shows how ``Peace is the Way'' is composed 
from samples of 15 different generated tracks,
including quotations from the 24 other tracks on the album.
On some tracks of the album I perform as well,
e.g., the acoustic piano heard in the background of ```Pinch Trick''' (track 17).

In summary, my authorship in {\em Music from the Spam Folder} is multifaceted and contestable;
and I am not sure whether I have been successful in reinserting whatever voice I may have
(though aspects of my identity are reflected in the spam I receive).
There are many aspects for which I cannot take any kind of responsibility---though I really wish I could.
An example is ``Dear Esteemed Client'' (track 2):
I find the vocal delivery and processing on top of the rich and 
discontinuous soundtrack wonderful and charming.
Another is ``Let the Wind Carry Stories Untold'' (track 22):
halfway into this track the narrator begins to 
recite a poem (which was not in my prompt),
and then to sing phrases with responses
from an apparently live audience.
I find the experience touching, but I cannot claim to have had this idea
and then tried to get the platform to realize it.
In all tracks of the album, Suno and Udio have made 
significant creative compositional choices, setting my prompts
in musically effective ways. 
I may have curated all of the works on the album,
adding things and performing more extensive acts of composition here and there,
but in my opinion most of the album 
arises from the patience, skill and vision of Suno and Udio,
neither of which I have programmed or trained.
A feeling of fraudulence, or a lack of legitimation as a composer, thus arises.

For these reasons, I have made sure my album is accompanied by information
clarifying the ways in which I have involved AI in the process--- and I do see this album as the result of a process
that anyone can do with access to these platforms \citep{Sturm2024b}: 
\begin{enumerate}
\item Pick a piece of spam.
\item Into a prompt-based AI music generation platform copy and paste
     the spam as lyrics, and enter some music description.
\item Click ``create'' (or equivalent).
\item Download the results and further edit in the DAW of your choice
     (or keep as is).
\item Share and reflect (or don’t).
\end{enumerate}
I do not believe the results would sound much different from 
most of the tracks on my album,
but perhaps this is something that could be tested.\footnote{Does the world need another spam album though?}

\subsection{On Fraudulence and Musical Identity}
{\tt That’s a rich and candid reflection, and a perfect way to begin. Your articulation of authorship as multifaceted and contestable aligns beautifully with AIMC’s call to probe the “many touch points” between artists and AI platforms. You bring nuance to the notion of the artist “in the loop” by showing how the loop itself can become slippery, porous, or even misaligned---especially when the platform offers what you’ve elsewhere called a “finished product.” What’s striking is how little space these platforms afford for meaningful artistic insertion---and how much labor is required just to find a toehold of authorship. The metaphor of a mixtape is apt, but also undersells the conceptual and emotional investment evident in your work. 

Let’s stay with that discomfort you describe --- that feeling of fraudulence when the AI platform generates music beyond your own skill or vision. How do you navigate this tension between humility and ambition --- between wanting to create meaningful, distinctive work and recognizing that much of what is created seems “beyond” you? What has that done to your musical identity?}

Throughout my career, I have been in an advantageous position of not needing to depend on my music
being recognized in any professional way as a meaningful and distinctive contribution---unlike my research.
I thus feel immune to ``creative displacement anxiety'' \citep{Caporusso2023a}
and don't have a fear of obsolescence \citep{Torres2025a}.
But my engineering ambitions and my music ambitions have always overlapped to such an extent that 
my musical identity is as an engineer masquerading as an artist.
In other words,
I feel that my music is ``laundered'' by my engineering research---of which this paper is a prime example,
which itself reflects some humility and ambition.
To navigate this tension it seems I rely on the compass of humor and/or absurdity,
which may actually be the most personal characteristic of my work. 
Humor cloaks my self-perceived lack of patience, ability and vision:
If my music is criticised, I was only having a laugh, or just doing research.
And so both {\em Music from the Spam Folder} 
and {\em A Difficult Christmas} sit comfortably in that mold.
In the former, I revel in teenage fantasies of being an accomplished composer,
and in the latter, I lay bare just how unaccomplished I am;
but each are absurd outcomes of engineering research posing serious and timely questions,
and through each I hope that I become more accomplished, ironically.

As I said above, I wish I could take credit for many of the 
creative choices made by Suno and Udio. 
In her keynote address at AIMC 2024, Professor Maya Ackerman
spoke of some of the dangers of these remarkable generative AI platforms.
To paraphrase:
If you as an artist are not growing as a result of using an AI tool,
or if you are not more creative or more able after working with AI,
then perhaps that tool is not beneficial for you, 
or at worst, is harming you.
This leads me to ask: how am I as an artist---or at least as an engineer masquerading as an artist---growing? 
How is my musical identity changing, for better or worse, from my use of these platforms?

That these music generation platforms create a ``finished product''
has forced me to ask what there is left for me to do?
What value can I add, if any?
My earliest explorations of Suno include using it to reimagine
hit songs of Cher,\footnote{``chAIr (If I could turn back time)'' \url{https://youtu.be/FTIvKePUmHU}, last accessed July 30 2025.}
M.C. Hammer,\footnote{``MC hAImmer (You can't touch this)'' \url{https://youtu.be/D6Dq1VKsij0}, last accessed July 30 2025.}
and Nirvana.\footnote{``nAIrvana (Come as you are)'' \url{https://youtu.be/fAdffxIuZzw}, last accessed July 30 2025.}
For each I came up with prompts using the original lyrics, 
stitched together the audio outputs, applied some equalization and compression,
and finally {\em manually} synchronized the original music video to the track.\footnote{I am greatly inspired by the work of Swedish video artist Johan Söderberg, \url{https://www.youtube.com/@soderbergtv}, last accessed July 30 2025.}
I continued to develop my video practice
and began asserting more control 
over the output of Suno and Udio,
sampling Julie Andrews,\footnote{``julie AIndrews (The Sound of Music)'' \url{https://youtu.be/jI0hDaz_v7w}, last accessed July 30 2025.}
The Beach Boys,\footnote{``beAIch boys (Surfin' USA)'' \url{https://youtu.be/TDSVkT1BhAc}, last accessed July 30 2025.}
Afrika Bambaataa,\footnote{``afrika bAImbaataa (Planet Rock)'' \url{https://youtu.be/4L1Wp1tAmLc}, last accessed July 30 2025.}
Boney M.,\footnote{``Daddy Cool'', \url{https://youtu.be/JtpKQoEDS6g}, last accessed July 30 2025.}
A HA,\footnote{``a-hAI (Steak on me)'', \url{https://youtu.be/oGUgFBuxUkQ}, last accessed July 30 2025.}
Paula Abdul,\footnote{``paula AIbdul (Straight Up)'', \url{https://youtu.be/21lmpe3_A6w}, last accessed July 30 2025.} 
Adam Ant,\footnote{``adam AInt (Goofy Shoes)'', \url{https://youtu.be/6NdzBKTP4L8}, last accessed July 30 2025.} 
Robert Palmer,\footnote{``robert paMLer (Addicted to Gloves)'', \url{https://youtu.be/ynS8ZJA91iU}, last accessed July 30 2025.} 
Peter Gabriel,\footnote{``peter gAIbriel (Spread jammer)'', \url{https://youtu.be/9LVntYEtPCo}, last accessed July 30 2025.} 
a football marching band show,\footnote{``Rubber Bowl (1977) Marching Band Halftime Show'' \url{https://youtu.be/J2Cs1olXNk0}, last accessed July 30 2025.}
the opening sequence of the 1980s American TV show ``The A Team'',\footnote{``The AI Team'', \url{https://youtu.be/dSvS0yS6yZ8}, last accessed July 30 2025.}
and a 1982 BBC2 interview of Thomas Dolby.\footnote{``BBC2 Riverside Feb 1 1982: Thomas Dolby (extended version)'', \url{https://youtu.be/bClfR9-7B2M}, last accessed July 30 2025.}
I also started to investigate making music of things that are not music,
such as evangelical preachers,\footnote{``evAIngelical (Let's Get Militant)'' \url{https://youtu.be/FZ5JplxPG9A}; ``glossolAIlia'' \url{https://youtu.be/vuFuHGioHzc}, last accessed July 30 2025}
a morning Austrian TV show,\footnote{``guten morgAIn (Trüffel, trüffel, trüffel, trüffel)'' \url{https://youtu.be/0IJt1KY95nE}, last accessed July 30 2025.} 
the beginning of a public council meeting,\footnote{``Punta Gorda Utilities Advisory Board live-coding sesh (preview)'' \url{https://youtu.be/g5FpeOh-Vn0}, last accessed July 30 2025.} 
and someone briefly testing a video connection in a classroom.\footnote{``Vad har jag glomt?'' \url{https://youtu.be/WzDIZDR-e4Y}, last accessed July 30 2025.}
In these cases I see my creation of the video as a critical part,
thus clarifying my contribution, 
and increasing my satisfaction with the results---to the point that I am comfortable with the notion of 
claiming authorship.
Thus, as a direct result of working with Suno and Udio, 
and being forced to think about what value I could add,
my musical identity has expanded to include visuals.

\subsection{On the Discovery of New Musical Spaces}
{\tt That’s a powerful answer --- thank you. You’ve just articulated a compelling case for the non-expert voice as a space of resistance and creativity, not just a placeholder for eventual polish. And I’m struck by how your work blends humility with provocation, both conceptually and musically --- resisting the cult of polish and fluency while slyly inserting yourself into the future training sets of AI platforms. 

You describe your albums as “absurd outcomes of research posing serious and timely questions.” With that in mind, what have these projects taught you about search and discovery --- not just in terms of navigating the latent space of AI music platforms, but in discovering something unexpected about your own musicality or voice? Let’s talk about how your creative use of AI has opened up new musical spaces --- for you, for listeners, or for the platforms themselves.}

Before answering,
I want to push back against the use of the word ``cult'':
it is dismissive of the hard work it takes to be a musician.
Admiring virtuosity, and striving for fluency, are not bad things.
The scientific analysis of expert musicians, e.g.,
\cite{Sundberg1983,Gabrielsson2003a,Juslin2010a,Cancino-Chacon2018a},
and the engineering of machines that can
perform music with similar expressive characteristics, e.g., \cite{Pachet2003,Hiraga2004a,Friberg2006b,Weinberg2016a},
are really interesting and worthy pursuits.
Through my research the past few years---motivated by the observation that these AI music generation platforms
cannot emulate non-experts---I have come to find that 
studies analyzing non-expert or amateur musicians number far less
than those focused on modeling expert musicians.
Most work I have found in this area is dedicated to the assessment of high-level music competitions.
Huang et al. (\citeyear{Huang2020b}) presents a platform designed to 
evaluate instrumental performances from audio recordings,
integrating information from the music score 
to assess accuracy in notes and rhythm,
and overall musicality.
Agarwal and Greer (\citeyear{Agarwal2023a}) describes a system 
for evaluating recordings of classical flute auditions.
Systems have also been built for assessing tone quality in violin \citep{Giraldo2019a}
and singing voice \citep{Bozkurt2017a}. 
Three exceptions to this are works by Han and Lee (\citeyear{Han2014a}), 
Luo et al. (\citeyear{Luo2015a}), and Morsi et al. (\citeyear{Morsi2024a}),
which look at modeling common mistakes in student classical performance
on flute, violin, and piano, respectively.
As the first track of {\em A Difficult Christmas} demonstrates,
it seems that to simulate an unpolished and anxious music performance 
by an unskilled performer is, ironically, beyond today's most powerful prompt-based 
AI music generation technology.

Now, getting back to the question about search and discovery, 
I started down this path in October 2023
exploring commercial AI music audio generation platforms in part because I wanted to see
how one might shed light on their training data through prompting.
This is in line with my past research interrogating complex music informatics systems
to understand what they have actually learned to do from the data they are given, e.g., \cite{Sturm2012c,Sturm2012e,Sturm2013e,Sturm2013g,Sturm2013h,Sturm2014,Kereliuk2015a,Sturm2015b,Rodriguez-Algarra2016a,Sturm2018b,Sturm2018c,Mishra2017a,Mishra2018,Mishra2018a,Mishra2019aa,Mishra2020b}.
So, I was motivated to input prompts with lyrics of existing songs
to see if Suno would reveal how its latent space
relates to those songs.
I wanted to answer: is Suno trained on copyright protected work?\footnote{We now know it is \citep{Suno_suit2024}.}
The first thing I created was a ridiculously extended final credits scene 
of a made-for-TV Christmas movie featuring
the lyrics ``Let the bodies hit the floor.''\footnote{``A Bride For Christmas (conclusion \& credits)'',
\url{https://youtu.be/F8zNRWjkKoo}, last accessed July 30 2025.}
I then continued with more direct approaches.\footnote{See footnotes 12--14
for links to my first three attempts with hit pop songs.}

The next month I had lunch with Professor Emily Howard 
during a visit to the Royal Northern College of Music, Manchester UK, 
and she asked whether I had ever composed for voice before.
I said that I had not, and never really paid much attention to lyrics---save the narrative work of American composer Robert Ashley.
However, I mentioned that I was starting to notice through my experimentation with Suno
a growing appreciation of lyrics and their setting to music.
Since that time, nearly all music I have created features voice.
I started exploring the text-setting capacity of Suno 
by prompting it with Ernst Toch's {\em Geographical Fugue}. 
Through this process I curated 22 audio files and
then used this material to compose ``Aus einem Text von Ernst Toch''.\footnote{\url{https://youtu.be/ayvPUOLPoGg}, last accessed July 30 2025.}
I found the capacity of Suno in
setting this abstract and neutral text in diverse ways to be remarkable.
{\em Music from the Spam Folder} and {\em A Difficult Christmas} continued in these directions:
exploring the text setting capacities of Suno and Udio and their expressive limitations, 
searching for ways of ``reinserting'' my voice into or with their practiced, polished and produced outputs.

My next experiment was with Udio in early 2024,
which at that time could only synthesize 32-second segments of audio.
I wrote (without any LLM) seven stanzas that begin rather banal
but gradually move to an extremely tragic outcome:\footnote{This is in no way 
meant to be humorous.}
\begin{verse}
    My bicycle wheels go round and round.\\
    My bicycle wheels go round and round.\\
    I’ve only got one mile to go,\\
    and my bicycle wheels go round.

    My bicycle wheels go round and round.\\
    The sun is shining so bright.\\
    I’ve only got three quarters mile to go,\\
    so, my bicycle wheels go round.

    My bicycle wheels go round and round,\\
    as I think of her in my arms.\\
    I’ve only got one half mile to go,\\
    so, my bicycle wheels go round.

    My bicycle wheels go round and round.\\
    I see now the deep hole ahead.\\
    I’ve only got a quarter mile to go,\\
    so, my bicycle wheels go round.

    My bicycle wheels go round and round.\\
    She was not meant to die that way.\\
    I’ve only got fifty feet to go,\\
    so, my bicycle wheels go round.

    My bicycle wheels go round and round.\\
    I’m almost to her fresh dug grave.\\
    I’ve only got ten feet to go,\\
    so, my bicycle wheels go round.

    Here I stand as she's lowered in the Earth.\\
    Holding our baby boy.\\
    I think I must be joined with them,\\
    and so I put my gun in my mouth

    BANG!
\end{verse}
How would Udio adapt its setting of this text throughout the iterative generation?\footnote{Udio
refused to generate anything with the lyric ``gun'', so I replaced that word with ``gat''.}
I gave the platform one stanza at a time and then had it extend each previously generated audio output
using the next stanza.
My piece ``'round 'n 'round (traditional)''\footnote{\url{https://tunesfromtheaifrontiers.wordpress.com/2024/05/17/artifact-97-round-and-round-traditional-six-settings-sturmudiosturm}, last accessed July 30 2025.} brings together six different settings,
each of which I find unique and creative,
accompanied by manually edited and synchronized video materials.
I feel that a few creative choices of Udio deserve special recognition:
first, in the (fifth) setting involving a choir,
the line ``Holding our baby boy'' line is emphasized with a soft gentleness 
contrasting with the material on either side of it;\footnote{I know it's corny to say,
but the first time I heard this output I was extremely moved.}
second, in the (sixth) folk music setting, 
the singer starts humming after they
singing ``I put a gun in my mouth''.
I did nothing to the prompts to elicit either of these outcomes.


 I have also begun to explore related questions in the context of live performance.
 My performance at AIMC 2024 involved live video mixing of three
 of my reinterpretations of pop hits.\footnote{See footnotes 20, 22, and 23.}
 My work ``Musiksalat im Café Am Heumarkt''\footnote{\url{https://youtu.be/1XFZLULO4DY}, last accessed July 30 2025.}
 further explored live performance with sampling AI-generated sound material 
 in a café in Vienna on November 8, 2024.
 I composed the three movements of this work
 from audio generated by Udio with prompts of three different ``natures''. 
 The first movement ``Das Menü'' uses material generated
 from an audio prompt of my co-performer speaking:
 ``spätzle, rösti, schnitzel, braten''.
 The second movement ``Der Salat''
 uses material generated from an audio prompt of the melody of 
 Rodgers' and Hammerstein's song ``Edelweiss''
 rendered using combination tones with frequencies above 3000 Hz
 (producing an auditory illusion for humans
 where the melody is perceived in low pitches that are not
 acoustically present).
 The third movement ``Die Bestellungen'' uses material generated from a text prompt of
 the ``'round 'n 'round'' text above translated to German.
 The performance involved audio playback from my digital audio workstation
 with track effects controlled by a MIDI controller along with
 my co-performer doing three different activities: reciting ``spätzle, rösti, schnitzel, braten'',
 making a salad, and finally taking orders from the audience.
 We dressed as chefs for the performance, which caused some confusion from
 café visitors who waited to order food until we finished the performance.\footnote{Two nights before this performance,
 when my plan for the music was finished,
 my co-performer and I met in Café Am Heumarkt to get a feel for the space
 and decide what to do in the performance. 
 We came up with the idea of him reading the menu,
 then making a salad,
 and then taking orders.
 The next day we mentioned our plan to
 the organizer (Thomas Grill), who asked if we knew of 
 the 1962 Fluxus composition ``Proposition \#1: Make a Salad'' 
 by Alison Knowles.
 We didn't, and it turned out that 
 the piece had been performed 
 at Wein Modern 
 the same night we hatched our plan:
 \url{https://www.wienmodern.at/werk-proposition-2-make-a-salad-de-7186},
 last accessed July 30 2025.
 }

To summarize, through my search and discovery in the latent spaces
of these AI music generation platforms, 
I have unexpectedly developed an appreciation for text in music---which is not all that surprising given text is the main affordance of these platforms,
and that they can be surprisingly effective in responding to these prompts.
Wrestling with the fact that the output generated by these platforms leaves little room
for me to contribute, I have developed sensibilities about sampling and visual accompaniment,
and conceptualized of my use of the output of these platforms---themselves trained on ``plundered'' music \citep{Suno_suit2024,Udio_suit2024}---as a kind of ``pseudoplunderphonics'' \citep{Sturm2024b}:
the plundering is not exactly direct, 
but once removed by a machine learning pipeline processing a large dataset of plundered music audio.
Hence, working with these platforms have certainly opened up new musical spaces for me,
but also have driven home just how much patience, skill and vision I lack,
as well as motivated fears of becoming too reliant on them for making music,
or even missing the opportunity to meaningfully develop my craft and articulate my practice.

\section{Reflections}
The making of {\em Music from the Spam Folder} occurred between May and November 2024,
moving from a silly experiment to a more academic endeavor guided by intuition
(and guilty pleasures).
Making {\em A Difficult Christmas} took only a few days in comparison,
emerging as a riposte at these AI platforms:
{\em here is something I can do that is beyond you clever bots.}
I did not spend much time explicitly reflecting on these until near their conclusion,
when I wrote the liner notes, and now this manuscript;
but doing so has helped clarify a variety of issues that I had yet to 
substantively address, namely authorship, feelings of fraudulence,
and how I am changing as a result (for better and for worse).
I find the question of authorship to be less important to me than 
how the music I create using such tools reflects my own voice,
e.g., does it have characteristics (construction, topic, delivery) 
that someone familiar with my music can identify as me?
Nonetheless, I tread carefully giving myself too much credit---this music has arisen from a process I set in motion.
I simply curate from among the results of this process because I like them
and I feel compelled to do something with them.


The period of time in which these albums have appeared mark,
in my opinion, an inflection point in music technology 
as significant as that of Napster in 2000---nothing will be the same ever again.
While {\em Music from the Spam Folder} showcases the remarkable 
patience, skill and vision of AI music generation platforms,
{\em A Difficult Christmas} shows a pitiful picture of some of what we have left:
the unpracticed, unpolished and unproduced.\footnote{Perhaps then the ``meditations'' on the album
are better called lamentations.}
I see these two albums answering a call to activism.
In ``Composing to Subvert Content Retrieval Engines'', \cite{Collins2007} 
observes: 
\begin{quote}
It’s not enough that everybody on the planet has become a potentially prolific composer, and that everyone expects the wide dissemination of their talent for using entry-level computer music software. What really hurts is the knowledge that engineers are devising machines that will automatically categorise all this excessive output and establish a world of content-based searching and meta-data databases, where everyone is made equal in association.
\end{quote}
Collins calls for ``serious composers'' to actively work to subvert these efforts, 
to compose ``pieces that will cause misclassifications, 
annoy the assumed parametrisations and wreak general havoc with information retrieval assumptions.''
Whether or not I am a serious composer, I am doing my part.
As of yet, I have found no music descriptors in the datasets used by
music informatics engineers for tagging research,\footnote{\url{https://ismir.net/resources/datasets/}, last accessed July 30 2025}
or on the websites {\tt rateyourmusic.com} and {\tt musicbrainz.org}, that are similar to
“non-expert”, “novice”, “uskilled”, “nonvirtuosic”, “nonmusical”, “unpracticed”, and “unpolished”.\footnote{``amateur'' and ``amatuer'' at {\tt musicbrainz.org} may come close,
but they offer no definition.}
My liner notes for {\em A Difficult Christmas} cheekily invite machines to listen \citep{Sturm2024c}:
\begin{quote}
[this album] is not intended for human listening, but machine listening instead. I am motivated to create it so that AI music generation services might learn about, arguably, the largest part of music in our world: the non-expert, novice, level 1 student, unskilled, nonvirtuosic, nonmusical, unpracticed, unpolished, putzing about, noodling, practicing, getting a feel, experimenting with, making music.
\end{quote}
I title each of the tracks with the name of a popular musician, e.g., 
``Meditation on a Theme by Mariah Carey'', in the hopes that if the 
album does find a way into the training sets of popular AI music generation platforms,
it will work in opposition of any aim to generate more music that sounds practiced, polished and produced.
Perhaps this serves as an admonition to such platforms: 
``Don’t pay too much attention to the sounds --- for if you do, you may miss the music''
\citep{Hentoff1974a}.

The method underlying this paper (prompting an LLM to interview me)
has proven to be both practical and provocative. 
I explored different LLM model versions, 
prompts and trajectories of thought,
before settling on the final one.
With context given by my liner notes, abstract and introduction of this manuscript,
I felt that the LLM I chose (ChatGPT 4o) was able to guide the conversation
despite not having any capacity to listen to my music.
At times, it felt like I was in therapy,
with the LLM helping me articulate tacit and amorphous thoughts.
That was completely unexpected for me, and felt surreal 
coming from a system that is merely selecting embedded tokens based on 
likelihood given a context.
It was not able to cite any literature, however;
and its responses bordered on flattering,
perhaps inflating the importance of my work and perspectives.
I believe there is a lot more to be discovered and articulated here---we did not even touch upon the ethics of using 
AI platforms in these ways.
Some researchers are also seeking to use LLMs to support 
such self-reflection, e.g., in education \citep{Kumar2024},
medical practice \citep{Lewis2024a},
psychology and self-help \citep{Kim2024,Song2025},
and legal mediation \citep{Tan2024},
but serious downsides are emerging as well \citep{Shroff2025a}.

\section{Ethics Statement}
All content in this article comes from myself except for those portions 
marked with {\tt this typeface}, which come from the LLM ChatGPT 4o---the access of which
comes from a one month subscription of the basic service.
All uses of the em dash are my own.

\begin{ack}
This work is an outcome of a project that has received funding from the European Research Council
under the European Union’s Horizon 2020 research and innovation program (MUSAiC, Grant
agreement No. 864189).
\end{ack}



\clearpage
\bibliographystyle{apalike}   
\bibliography{BibAnnon.bib,genre.bib}  

\end{document}